\documentclass[prd,floatfix,twocolumn,preprintnumbers,letterpaper]{revtex4}
\usepackage{graphicx}
\usepackage{amsmath,amssymb}
\usepackage{epsfig}
\usepackage{latexsym}
\newcommand {\ga} {\ {\raise-.5ex\hbox{$\buildrel>\over\sim$}}\ }
\newcommand {\la} {\ {\raise-.5ex\hbox{$\buildrel<\over\sim$}}\ }

\begin{document}

\def\be{\begin{equation}}
\def\ee{\end{equation}}

\title{The Late Time Behavior of False Vacuum Decay:  Possible Implications for Cosmology and Metastable Inflating States }
\author{Lawrence M. Krauss$^{1,2}$ and James Dent$^{2}$}
\affiliation{$^1$CERCA, Department of Physics, Case Western Reserve University,
Cleveland, OH~~44106}
\affiliation{$^2$Department of Physics \& Astronomy, Vanderbilt University,
Nashville, TN~~37235}
\date{\today}

\begin{abstract}
We describe here how the late time behavior of the decaying states, which is predicted to deviate from an exponential form, while normally of insignificant consequence, may have important cosmological implications in the case of false vacuum decay.  It may increase the likelihood of eternal inflation, and may help explain the likelihood of observing a small vacuum energy at late times, as well as arguing against decay into a large negative energy (anti-de Sitter space), vacuum state as has been motivated by some string theory considerations.  Several interesting open questions are raised, including whether observing the cosmological configuration of a metastable universe can constrain its inferred lifetime.
\end{abstract}

\maketitle

\section{Introduction}
It is not commonly known, but nevertheless well established since the work of Khalfin in 1958\cite{khalfin},  extended by numerous other authors in the intervening years (i.e. \cite{sudarshan,peres,patra} that for times long compared to the characteristic decay time of a metastable quantum state, the decay of such states is no longer described by an exponential, but rather by a power law.  

One of the reasons that this interesting fact is perhaps not more well recognized is that the characteristic time-scale determined for departures from exponential decay in several sample systems is generally so long that for all intents and purposes all unstable states in any finite system will have decayed before this onset would become relevant.  

However, if one generalizes the results from quantum mechanics to quantum field theory then the nature of false vacuum decay can altered, with potentially significant implications for cosmology, as we describe here.

We begin by exploring the generic features of late-time decay, examining the transition from quantum mechanics to quantum field theory in order to explore those implications that are independent of the details of the relevant potential or model, in order to consider their impact upon the dynamics of cosmological false vacuum decay.  We find that the time for departures from exponential growth should depend logarithmically on ratio of the energy gap and the decay width with a prefactor of order unity and then explore the cosmological implications of this.    Several open questions are then raised, including whether gravitational effects may alter the long time non-exponential tail in vacuum decay, and also whether cosmological observation within a given metastable universe may alter its lifetime.

\section{Late Time Decay in Quantum Mechanics}

In 1958 Khalfin \cite{khalfin} gave a general argument for why the exponential decay of an unstable state  breaks down at late times by employing the Payley-Wiener theorem \cite{paley} to show that if the energy distribution function is bounded below, then its Fourier transform, the survival amplitude, necessarily leads to a power-law decay.  (Notably the Breit-Wigner form of the energy distribution function is not bounded and does lead to exponential decay).  A key question of course is when the deviation from exponential decay will begin.  While this is in general a model-dependent question, it turns out, as we demonstrate by utilizing three different \cite{sudarshan,peres,patra}, but particularly useful formalisms of deriving the power law decay at large times, that the overall time-scale has various  model-independent features.  

First, following \cite{sudarshan} we consider an initial unstable state, $|M>$, for which the survival probability $P(t)$ is given by
\begin{eqnarray}
P(t) = |<M|e^{-iHt}|M>|^2
\end{eqnarray}
where $H$ is the Hamiltonian.  $P(t)$ is equal to the square of the survival amplitude $a(t)$ which may be written as the Fourier transform of the energy distribution function, $\rho(E)$
\begin{eqnarray}
a(t) = \int e^{-iEt}\rho(E)dE
\end{eqnarray}
Here we will assume that the energy distribution function is bounded below so that there exists a lowest $E'$ such that $\rho(E) = 0$ for $E < E'$ and thus the Paley-Wiener theorem applies which will insure a non-exponential decay at large times.

The survival amplitude $a(t)$ can be represented using a contour integral in the complex $z$-plane (see figure 1):
\begin{eqnarray}
a(t) = <M|e^{-iHt}|M> = \frac{1}{2\pi i}\int_{C}e^{-izt}\beta(z)dz
\end{eqnarray}
where $\beta$ is given in terms of the resolvent $(H-z)^{-1}$ as
$\beta(z) = <M|\frac{1}{H-z}|M>$.

There is a cut along the spectrum of $H$ which is taken for simplicity to be given by $(0,\infty)$ along the real axis. Upon analytic continuation to the second sheet, $\beta(z)$, which  is analytic in the cut plane and has no zeros there, contains poles  which give rise to the exponential decay law as well as corrections which lead to the power law decay at large time.   The resonance condition requires a  pole at $z = E_o -\frac{i}{2}\Gamma$, with $E_o \gg \Gamma$.  Physically $E_o$ is the energy difference between the unstable state and the state into which it decays. 

\begin{figure}
  \begin{center}
    \epsfig{file=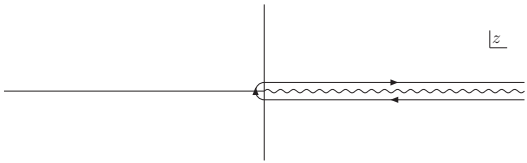,height=27mm}
  \end{center}
  \caption{Contour for $a(t)$ integration
  \label{fig1}}
\end{figure}

In order to relate this expression for the survival amplitude to previous studies, it is simpler to work with the inverse of $\beta(z)$, $\gamma(z) = 1/\beta(z)$.  $\gamma(z)$ may be written in terms of dispersion relations whose exact form depends on the asymptotic behavior of $\gamma$.  

It is useful to frame this problem in terms of a formalism that will later be useful for thinking about field theory.  One thus can consider a 'density of states' factor $f(E)$ defined in terms of $\gamma$:
\begin{eqnarray}
|\tilde{f}(E)|^2\sigma(E)= \frac{1}{2i}\lim_{\epsilon \rightarrow 0}(\gamma(E + i\epsilon) - \gamma(E-i\epsilon)) 
\end{eqnarray}
where $\sigma(E)$ is a phase-space weight factor.  
Now $a(t)$ may be written as
\begin{eqnarray}
a(t) &=& \frac{1}{\pi}\int_{0}^{\infty}\frac{|\tilde{f}(E)|^2}{|\gamma(E+i\epsilon)|^2}\sigma(E)e^{-iEt}dE\\
&\sim& \frac{1}{\pi}\int_{0}^{\infty}\sigma(E)e^{-iEt}dE
\end{eqnarray}
Then using that $\sigma(E) \sim E^{1/2}$ (which is also true in the relativistic case)
\begin{eqnarray}
\int_{0}^{\infty}e^{-iEt}E^{1/2}dE = t^{-3/2}\int_{0}^{\infty}e^{-iut}u^{1/2}du
\end{eqnarray}
one can see that $a(t) \sim t^{-3/2}$ at large times (and the survival probability goes as $t^{-3}$).

To return to the key issue of when the turnaround between exponential and power-law behavior, one must be more specific in calculating $a(t)$.  A variable change  \cite{sudarshan}  simplifies  the $a(t)$ integration.

Defining
$k = z^{1/2}e^{i\pi/4}$, then
due to the resonance condition and the jump discontinuity across the branch cut, $\gamma(z)$  has poles on the second sheet at $z=E_o \pm \frac{i}{2}\Gamma$ (where $E_o \gg \Gamma$) which, in terms of $k$ occur approximately at
$k_{\pm} \simeq \pm k_o + \delta$
where
\begin{eqnarray}
k_o = E_{o}^{1/2}e^{i\pi/4}\,\,\,;\,\,\, \delta = \frac{\Gamma}{4E_{o}^{1/2}}e^{-i\pi/4}
\end{eqnarray}
One can then show the zero structure of $\gamma(z)$ (or the pole structure of $a(t)$) more clearly as
$\gamma(k) = e^{-i\pi/2}(k-k_+)(k-k_-)\xi(k)$, 
which places all model dependence into the function  $\xi(k)$.  Using this, $a(t)$ can now be written as a sum of an exponential decay piece
as well as a 'background integral' term.
In \cite{sudarshan} two specific models for $\xi(k)$ were studied, and interestingly in both cases the crossover time was precisely the same, and given by:

\begin{eqnarray}
\label{large time}
T \sim \frac{5}{\Gamma}ln\bigg(\frac{E_o}{\Gamma}\bigg) + \frac{3}{\Gamma}ln\bigg(5ln\bigg(\frac{E_o}{\Gamma}\bigg)\bigg).
\end{eqnarray}      

The fact that the logarithm of the ratio $\frac{E_o}{\Gamma}$ appears in this result is not surprising, as these are the only two dimensional parameters that appear in the general problem and the crossover time will be based on a comparison of the magnitude of an exponential decay rate and a power law decay rate.   The question that remains is to what extent the pre-factor arises from model-dependent features, and how things may generalize in quantum field theory.   

In order to explore this, it is useful to examine the same result within other formalisms, .  In \cite{peres} a Hamiltonian approach was taken where one writes the full Hamiltonian as a sum of terms, and considers a basis of incoming states that diagonalize part of the Hamiltonian, treating the remainder as a perturbation.  In this case by expanding a wavefunction in this diagonal basis one can calculate the time dependence of the coefficients.  The time dependence is in turn given by the integral over matrix elements between the initial state and all states with nearby energy, with the time dependence of the amplitude being given by:
\begin{eqnarray}
\dot{c}_0(t) = -\int_{0}^{t}w(t-t')a_0(t')dt'
\end{eqnarray}
where 
\begin{eqnarray}
w(t) \equiv \int e^{-i(E-E_{in})t}dF(E)
\end{eqnarray}
Introducing the function $W(E)$ which is the Fourier transform of $w(t)$
\begin{eqnarray}
W(E) \equiv \frac{dF(E)}{dE} = \frac{1}{2\pi}\int_{-\infty}^{\infty}w(t)e^{i(E-E_{in})t}dt
\end{eqnarray}
 the survival amplitude of a state is controlled solely by the integral of the function $W(E)$.  
 
A specific example for $W(E)$ appropriate for a particle in three dimensions then allows an explicit calculation of the cross-over time \cite{peres}
which is comparable to that given in the previous analysis, and in particular again depends logarithmically  on $(E_o/\Gamma)$.  

\section{Semi-Classical Methods and False Vacuum Decay}

Presenting these results in this way makes the decay rate calculation reminiscent of a WKB calculation, where an integral over the action of the barrier between states plays a key role, and where semi-classical steepest descent estimates are often used in the calculation of false vacuum decay in field theory.  This approach, at least in the context of a simple quantum mechanical model, was used to calculate non-exponential decay at large times in \cite{patra}.  For the simple case of a particle trapped in a square potential well the Euclidean (imaginary time in Minkowski space) kernel was calculated
and at large times was shown to exhibit a power law behavior $\approx T^{-3/2}$,
which could then used to approximate the cross-over time with precisely the same prefactor as that given in (\ref{large time}).

With a Euclidean time formalism for calculating late time decay one can straightforwardly extrapolate to the case of false vacuum decay in field theory.  We now address various issues associated with this extrapolation.  

(i) The examples above involved single particles in  one or three dimensions and one may thus question whether the power law dependence derived there is appropriate in a field theoretical context.  To address this question one can make use of a heuristic analysis first applied to false vacuum decay \cite{guth}, which reproduces the above result and also illuminates the origin of the power law dependence in a way that directly applies to the field theory calculation.  

For tunneling from a barrier in one dimension there is an exact solution to the time-dependent Schrodinger equation, but it is not normalizable, and is therefore only an approximation to the true physical wavefunction.  A wavepacket originally centered at the local minimum of the potential energy will initially exponentially decay, but the outgoing wavepacket to the right will spread as it evolves.  Thus there will be some backflow of probability at the origin.  The spread will grow linearly with time $T$, and the density of the packet, extrapolating to $d$ dimensions will fall like $1/T^d$.   

In the field theory case one extrapolates to an infinite number of degrees of freedom.  However, as long as the euclidean bounce solution has only one negative eigenvalue \cite{coleman}, the number of effective degrees of freedom associated with the wave-packet spread effectively reduces to the single particle case, with the the appropriate dimensions in the late time behavior of decay being the dimensionality of space. 

(ii) The effect of going from quantum mechanics to field theory is to include, among other things, the need to consider volume factors so that one calculates survival probabilities per unit volume per unit time.   In the standard false vacuum decay calculation, for example, the fact that euclidean time solutions (instantons) can be placed throughout the full volume of space time is of importance.  However, since the same volume factors should appear in both the estimate of the early and late time decay rate estimates, it is reasonable to assume that the crossover time estimate derived above will apply in this case to survival probabilities per unit volume.   

In particular, recognizing that both the exponential and non-exponential decay rates depend upon the same integral and phase space factors as described above, we infer that the result in (\ref{large time}) should give a quantitative estimate of the crossover time in which false vacuum decay rate per unit volume converts from exponential to power law, namely:

\begin{eqnarray}
\label{largisher time}
T \sim \frac{O(5)}{\Gamma}ln\bigg(\frac{E_o}{\Gamma}\bigg) 
\end{eqnarray}   

This implies that the survival probability (per unit volume) at this time would be given by:
  
 \begin{eqnarray}
\label{largish time}
{N \over N_0} \approx \bigg(\frac{\Gamma}{E_o}\bigg)^{5}
\end{eqnarray}
 
We shall use this estimate in the discussions that follow. 

\section{Late TIme Decay of the False Vacuum and Implications for the landscape of Inflating Universes}

Typically for laboratory systems one finds that ${E_o}/{\Gamma} \ge O(10^3 -10^6)$ so that (\ref{largish time}) implies the exponential suppression in the number of decaying states is large, typically involving a reduction by $\approx O( exp(-30))$.   As a result, this phenomena is generally unobservable.   However, it is well known in inflationary cosmology that even exponentially suppressed processes can be significant.  For example, in eternal inflation, even though regions of false vacua are assumed to decay exponentially, gravitational effects cause space in a region that has not decayed to itself grow exponentially fast.  Thus most of space can, at any time, be inflating, even well after the canonical time in which one would consider, in the absence of gravity, the phase transition to have been completed.

The result we describe here appears to exacerbate this situation considerably.   Those false vacuum regions which do survive well beyond the canonical decay time, up to the time given in (\ref{largisher time}), may effectively never decay, as within these regions the survival probability per unit volume will fall as a power law, while the volume of the regions will grow exponentially.  In the long term, then, space will become dominated by precisely these regions. 

If this is the case, the most probable regions of space to find oneself within, even long after the canonical false vacuum decay time would be metastable regions.  Indeed, the fact that survival probability only falls as a power law at late times implies that one can have eternal inflation scenarios even when the decay rate is comparable or even somewhat larger than the inflationary expansion rate.  In such scenarios the volume of space associated with false vacuum will initially decrease, but at late times, when the power law kicks in, the volume of space in false vacua may quickly begin to increase again. 

Because the ratio of metastable energy to decay rate appears in the crossover time, the absolute survival probability at late times for any specific region of inflating false vacuum will increase as the vacuum energy decreases for a given fixed decay rate.  (Because, in the WKB approximation the decay rate depends primarily upon the integral of the action under the barrier it is not implausible that this decay rate can be a slowly varying function of the energy gap.)   While probabilities in eternal inflation scenarios are notoriously difficult to normalize, this argument suggests that the survival probability may be largest in those metastable regions with the smallest allowed vacuum energy.   

Is it therefore coincidence that we currently find ourselves in a universe with extremely small vacuum energy?  One cannot answer this question unambiguously without knowledge of the underlying theory but since exponential expansion has only begun in our universe relatively recently, if our vacuum is metastable we are not likely to exist in the late time regime.  Specifically, given the inferred vacuum energy today, if we assume the ultimate ground state vacuum energy is zero, then $E_0 \approx 10^{-2}$ eV.  If our vacuum were metastable, the time at which power law decay probability would set in would depend upon the unknown decay rate.  If the canonical decay rate for our vacuum is less than about twice the present Hubble constant, $O (SH_0) \approx 10^{-32} $ eV, so that vacuum decay would generically not occur until after exponential expansion began (and after finite temperature effects would presumably no longer stabilize the false vacuum),  the onset of late time power law decay will not occur until the universe is at least 30 times its current age.   (If the decay rate were much faster, then we would generically expect the fraction of space which would survive as a function of time would have been decreasing dramatically over cosmic time before the current vacuum energy began to dominate, which does not seem likely given the size of the currently observable universe.)

These considerations provide a small modicum of hope.   If our universe is metastable, the expected survival time may be not much larger than the current Hubble age.  However,  we might take solace in the possibility that if our region is one of the rare late-time survivors, vacuum decay will occur over an exponentially decreasing fraction of our space as time proceeds.

Questions of the long-term survival of universe aside, these arguments do suggest that maximizing the volume of space contained in metastable vacua at late times will favor those regions with the smallest vacuum energies.    Alternatively, it would appear that a late time transition to a true vacuum state with large negative energy is disfavored.   If, for example, the canonical decay rate were comparable to the present age of the universe, and the energy gap to the true vacuum were  $O(10^{19})$ GeV, the exponential suppression factor would exceed $\approx exp (-300)$ before a period of power law decay would set in.   This argument disfavor string landscape scenario where the ultimate vacuum state is anti-de Sitter and anthropic arguments are used to explain our present configuration.

\section{Conclusions and Open Questions}

The possibility that macroscopic quantum mechanical considerations may impact upon the very existence of our own universe at late times is an intriguing one.   As we have seen, at the very least they may require a dramatic alteration of probability considerations in chaotic eternal inflationary universe scenarios, and may explain why universes with unnaturally small vacuum energies may be favored at late times, even as these arguments would tend to disfavor ultimate decay into a state with large negative vacuum energy.   

However, these considerations also lead us into unexplored realms associated with macroscopic quantum mechanics.     The first, which we plan to investigate, involves the question of whether late-time power law time dependence persists when gravitational effects in vacuum decay \cite{coleman2} are taken into account.  The heuristic argument regarding spreading of the  wavefunction following barrier penetration, leaving a power law tail at the origin as the source of late time behavior suggests that the exponential expansion of space accompanying metastability may suppress this term and thus produce a faster falloff.     The second consideration is even more interesting.   Have we ensured, by measuring the existence dark energy in our own universe, that the quantum mechanical configuration of our own universe is such that  late time decay is not relevant?  Put another way, what can internal observations of the state of a metastable universe say about its longevity?

\acknowledgements 
L.M.K. is supported in part by the DOE and NASA, and thanks Vanderbilt University for hospitality while this work was begun.
We thank Tom Weiler for making us aware of power law late-time decay, and Alan Guth, Glenn Starkman, Tom Kephart and Robert Scherrer for useful conversations.

\end{document}